\documentclass[twocolumn,preprintnumbers,nofootinbib,prd,superscriptaddress,aps]{revtex4-2}

\usepackage[utf8]{inputenc} 
\usepackage{graphicx,amssymb,amsmath,amsthm,amsfonts,times}
\usepackage[linktocpage]{hyperref}
\usepackage[usenames]{color}
\usepackage{epstopdf}
\usepackage{textcomp}
\usepackage{bm}
\usepackage{latexsym}
\usepackage{rotating}
\usepackage{color}
\usepackage{longtable}
\usepackage{enumerate}
\usepackage{stmaryrd}
\usepackage[normalem]{ulem}
\usepackage{mathtools}
\usepackage{url}
\usepackage{multirow}
\usepackage{graphicx}
\usepackage[caption=false]{subfig}
\usepackage{verbatim}
\usepackage{soul,xcolor}
\usepackage{xspace}

\setstcolor{red}

\makeatletter
\DeclareRobustCommand\onedot{\futurelet\@let@token\@onedot}
\def\@onedot{\ifx\@let@token.\else.\null\fi\xspace}
 
\makeatother

\definecolor{coolblack}{rgb}{0.0, 0.18, 0.39}
\definecolor{darkred}{rgb}{0.5,0,0}
\definecolor{darkgreen}{rgb}{0,0.5,0}
\definecolor{darkblue}{rgb}{0,0,0.5}
\definecolor{lapislazuli}{rgb}{0.15, 0.38, 0.61}
\definecolor{venetianred}{rgb}{0.78, 0.03, 0.08}
\definecolor{bleudefrance}{rgb}{0.19, 0.55, 0.91}
\definecolor{dogwoodrose}{rgb}{0.84, 0.09, 0.41}
\definecolor{dogwoodrose}{rgb}{0.84, 0.09, 0.41}
\definecolor{darkorgane}{rgb}{1,0.549,0}
\hypersetup{colorlinks=true, citecolor=darkblue, linkcolor=darkblue, 
urlcolor = darkblue}
\definecolor{olive}{rgb}{0.5, 0.5, 0.0}

\newcommand{\ben}{\begin{enumerate}}
\newcommand{\een}{\end{enumerate}}

\def\be{\begin{equation}}
\def\ee{\end{equation}}
\newcommand{\beq}{\begin{eqnarray}}
\newcommand{\eeq}{\end{eqnarray}} 
\newcommand{\ba}{\begin{align}}
\newcommand{\ea}{\end{align}}
\newcommand{\Ricci}{\mathcal{R}}

\def\be{\begin{equation}}
\def\ee{\end{equation}}
	
\newcommand{\bea}{\begin{eqnarray}}
\newcommand{\eea}{\end{eqnarray}}

\begin{document}

\title{Piercing of a solitonic boson star by a black hole}
\author{Zhen Zhong}
\affiliation{CENTRA, Departamento de F\'{\i}sica, Instituto Superior T\'ecnico -- IST, Universidade de Lisboa -- UL,
Avenida Rovisco Pais 1, 1049-001 Lisboa, Portugal}
\author{Vitor Cardoso}
\affiliation{Niels Bohr International Academy, Niels Bohr Institute, Blegdamsvej 17, 2100 Copenhagen, Denmark}
\affiliation{CENTRA, Departamento de F\'{\i}sica, Instituto Superior T\'ecnico -- IST, Universidade de Lisboa -- UL,
Avenida Rovisco Pais 1, 1049-001 Lisboa, Portugal}
\author{Taishi Ikeda}
\affiliation{Dipartimento di Fisica, 
``Sapienza'' Universit\'{a} di Roma, Piazzale Aldo Moro 5, 00185, Roma, Italy}
\affiliation{Niels Bohr International Academy, Niels Bohr Institute, Blegdamsvej 17, 2100 Copenhagen, Denmark}
\author{Miguel Zilhão}
\affiliation{Departamento de Matem\'atica da Universidade de Aveiro and
  Centre for Research and Development in Mathematics and Applications (CIDMA),
  Campus de Santiago, 3810-183 Aveiro, Portugal}

\begin{abstract}
Recently, the piercing of a mini boson star by a black hole was studied, with tidal capture and the discovery of a ``gravitational atom'' being reported~\cite{Cardoso:2022vpj}. 
Building on this research, we extend the study by including a hexic solitonic potential and explore the piercing of a solitonic boson star by a black hole. Notably, the solitonic boson star can reach higher compactness, which one might expect could alter the dynamics in this context. Our findings suggest that even when the black hole's size approaches the test particle limit, the solitonic boson star is easily captured by the black hole due to an extreme tidal capture process. Regardless of the black hole initial mass and velocity, our results indicate that over 85\% of the boson star material is accreted. Thus, the self-interaction does not alter the qualitative behavior of the system.
\end{abstract}

\maketitle

\section{Introduction}
Based on a range of existing observational evidence, it is widely accepted that most of the universe is made of dark matter and energy, which gravitate but otherwise interact at most feebly with the Standard Model particles~\cite{Freese:2008cz,Navarro:1995iw,Clowe:2006eq,Bertone:2004pz}. Thus far, endeavors to pinpoint the nature and attributes of dark matter and incorporate it within a theoretical framework have proved futile, but these efforts will persist in the foreseeable future~\cite{Kahlhoefer:2017dnp,PerezdelosHeros:2020qyt}.

It is reasonable to expect that, in the same way that Standard Model particles come together to form stars and planets, dark matter particles also form self-gravitating structures, or ``dark stars'' of various types due to gravity, consequently constituting a considerable portion of astrophysical environments.
Dark stars -- if they exist and are not black holes (BHs) -- have so far gone undetected, but the advent of gravitational-wave (GW) astronomy has the potential to revolutionize our knowledge of the universe~\cite{Barack:2018yly,Cardoso:2019rvt,Giudice:2016zpa,Ellis:2017jgp,Cardoso:2022whc}.
Here, we will focus on a special type of dark matter (which might constitute the whole or just a fraction of the dark matter content in the cosmos) -- light scalar fields. The possibility that the dark matter detected in galaxies is composed of ultra-light scalar particles in a Bose-Einstein condensate was put forward some time ago~\cite{Hui:2021tkt}.
The significance of these candidates in cosmology arises from the fact that their de Broglie wavelength is of comparable magnitude to astrophysical scales, potentially alleviating some of the tension with observations~\cite{Schive:2014dra}. These new fundamental fields can indeed form self-gravitating structures, which in a purely General Relativistic context are known as boson stars (or Proca stars when the fundamental constituents are massive vectors)~\cite{Liebling:2012fv}.

Thus far, efforts to study dark stars have focused mostly on collisions of boson stars with similar sizes and masses~\cite{Palenzuela:2017kcg,Sanchis-Gual:2018oui,Sanchis-Gual:2020mzb,Evstafyeva:2022bpr}, but some scenarios involving larger mass ratios have also been studied~\cite{Bezares:2022obu, Siemonsen:2023hko}.
As previously studied in~\cite{Davies:2019wgi, Cardoso:2022vpj}, it is possible that dark matter stars are structures with a larger scale, in which case it is important to study what happens when BHs or compact objects cross such a medium, or its effect on a tight, GW-emitting binary. In particular, dynamical friction, accretion and emission of dark matter will affect the dynamics of the system and a precise knowledge of the process is necessary~\cite{Cardoso:2022whc,Eda:2013gg,Macedo:2013qea,Barausse:2014tra,Hannuksela:2018izj,Cardoso:2019rou,Baumann:2019ztm,Kavanagh:2020cfn,Annulli:2020lyc,Zwick:2021dlg,Vicente:2022ivh, Cardoso:2022vpj}.

Numerical relativity can yield precise results for the dynamics and GW emission details, when the length scales of different objects are similar~\cite{Palenzuela:2017kcg,Bustillo:2020syj,Bezares:2022obu,Cardoso:2014uka,Siemonsen:2023hko}. However, situations where the length scales differ substantially are challenging, and probe the limits of current infrastructure. We have recently studied highly diluted mini-boson stars (without self-interactions) as a model for dark matter cores in halos, and examined the behavior of small BHs as they traverse through such large boson star structures \cite{Cardoso:2022vpj}. We observed dynamical friction and tidal-induced capture which led to the accretion of the entire boson star, even when it was two orders of magnitude larger than the BH itself.

The addition of a repulsive self-interaction term introduces extra resistance to gravitational collapse and may modify drastically the relevant length scales~\cite{Lee:2017qve}. Moreover, the maximum mass of mini-boson stars $M_{\mathrm{max}} \approx M_{\mathrm{Planck}}^2 / \mu$ is significantly smaller than the Chandrasekhar mass $M_{\mathrm{Ch}} \approx M_{\mathrm{Planck}}^3 / \mu^2$ for bosonic particle candidates with typical masses, where $\mu$ is the mass of particle~\cite{Liebling:2012fv, Seidel:1990jh, Herdeiro:2022gzp}. Dark stars are anticipated to accrete mass from their surroundings, regardless of their initial mass. As a result of this accretion process, their mass can increase by as much as $10^7 M_{\odot}$ \cite{Freese:2017idy}. To extend the limit of the potential to astrophysical masses that are comparable to the Chandrasekhar mass, a self-interaction component has been incorporated into the potential to provide additional pressure to counteract gravitational collapse \cite{Mielke:1997re}. 

Our purpose here is to generalize Ref.~\cite{Cardoso:2022vpj} to solitonic boson stars (SBSs). As the simplest example of nontopological solitons, an SBS can even exist in the absence of gravity and is referred to as a Q-ball~\cite{Coleman:1985ki,Liebling:2012fv, Boskovic:2021nfs, Cardoso:2021ehg}.
Mini-boson stars attain stability through the balance between gravitational and repulsive pressure forces. On the other hand, the stability mechanism of SBSs differs, with a bubble-like structure emerging in the densest region of the parameter space. Stability arises from the accumulation of energy near the surface, engendering surface tension among distinct vacua \cite{Boskovic:2021nfs}.

Evidence also suggests that SBSs can describe dark matter cores in sub-halos.
A sub-halo is a smaller clump of dark matter that is gravitationally bound within a larger dark matter halo, and is created when smaller halos are accreted and tidally disrupted by larger ones~\cite{Jiang:2016yts}. 
A solitonic core can be considered as a special type of sub-halo that has a different profile and properties than other sub-halos \cite{Schive:2014hza}. However, the variability of the core-halo relation can make the discrepancy between soliton and sub-halo less strong for larger halos \cite{Ferreira:2020fam}. If dark matter is made up of ultralight bosons, it is possible for solitonic cores to form at the centers of dark matter halos~\cite{Hui:2016ltb}. It is worth noting that Ref.~\cite{Mocz:2017wlg, Schwabe:2016rze} provides evidence that the density profiles of different mergers of solitonic cores conform to the SBS profile in ultralight axion dark matter halos. 
In addition, although the well-known cusp-core problem can be overcome by introducing a quartic term in the self-interaction scalar potential \cite{Harko:2011xw, Deng:2018jjz}, the scaling relation between the dark matter halo radius and central density still contradicts the observations \cite{Gavrilik:2020mjy}. With an additional $\phi^6$-term in the potential, the problem could be resolved and the cores would have a non-trivial phase structure \cite{Gavrilik:2020mjy, Gavrilik:2021gtp}.
In this work, by focusing on SBSs, we aim to gain a comprehensive understanding of how solitonic cores behave in the presence of drifting perturbers.
We note, however, that for relativistic fuzzy dark matter models, the boson mass is approximately of the order of $10^{-22}$ eV, which would correspond to a solitonic core radius of order $1$ kpc \cite{Schive:2014dra}. Considering the largest known astrophysical black hole, \texttt{Tonantzintla 618}, its radius is of the order of $10^{-6}$ kpc \cite{Shemmer:2004ph}, resulting in a length ratio of $10^6$ which is impossible to resolve with our current approach. As a model for such systems, we will present the largest ratios that are feasible with our computational infrastructure, and as we will argue later, we do not expect our results to change significantly for larger length ratios that remain within one order of magnitude. Probing length scales much higher than these currently considered would necessitate a fundamentally different approach.

We use units where $G=c=\hbar=1$ throughout.

\section{Framework}

\subsection{Solitonic boson star}
We consider the Lagrangian density of a self-gravitating, complex scalar field $\Phi$ with a solitonic potential $V=V(|\Phi|^2)$
\be
\mathcal{L}_m =\frac{\mathcal{R}}{16\pi}-\left[g^{a b} \nabla_{a} \Phi \nabla_{b} \Phi^{*}+ V\right],\label{eq:LagrangianDensity}
\ee
where $g_{ab}$ is the metric of the spacetime, $\mathcal{R}$ is the Ricci scalar, $\Phi^*$ is the complex conjugate of the scalar field and $V$ is the potential
\be
V= \mu^2|\Phi|^2 \left(1 - 2\frac{|\Phi|^2}{\sigma^2}\right)^2\,.\label{eq:potential}
\ee
Here $\mu$ is the scalar field mass and $\sigma$ is a free parameter controlling the self-interaction.
The self-interaction potential is chosen to provide configurations that can exist even in flat spacetime~\cite{Coleman:1985ki,Boskovic:2021nfs,Cardoso:2021ehg}, and is a standard choice in the literature (e.g.~\cite{Bezares:2017mzk,Palenzuela:2017kcg}).
Variation of the corresponding action with respect to the metric $g^{ab}$ gives the equations of motion
\begin{align}
&\Ricci_{ab} - \frac{1}{2} \Ricci g_{ab} = 8\pi T_{ab}\,,\\
&g^{ab}\nabla_a\nabla_b\Phi = \Phi \frac{d V}{d |\Phi|^2}\,,
\end{align}
with energy-momentum tensor
\begin{equation}
T^{a b}=\nabla^{a} \Phi \nabla^{b} \Phi^{*}+ \nabla^{a} \Phi^{*} \nabla^{b} \Phi %
-g^{a b}\left(\nabla^{c} \Phi \nabla_{c} \Phi^{*}+ V \right)\,.
\end{equation}
Following Refs.~\cite{Cardoso:2022vpj,Liebling:2012fv,BECERRIL2007263}, we write down equilibrium equations with the general, spherical symmetric metric in Schwarzschild-like coordinates
\be
ds^2 = -\alpha^2 dt^2 + a^2 dr^2 + r^2 d\Omega^2_2\,,
\ee
where $\alpha=\alpha(r),\,a=a(r)$. In addition, to get the time-independent solution, we assume a harmonic ansatz
\be
\Phi = \phi(r) e^{i\omega t}\,.
\ee
Then, the Einstein-Klein-Gordon system can be written as three coupled ordinary differential equations
\begin{align*}
  \frac{2a'}{a}&=\frac{1-a^{2}}{r}+8 \pi r\left[\left(\frac{\omega^{2}}{\alpha^{2}}+ \frac{\mu^2 (\sigma^2 - 2\phi^2)^2}{\sigma^4} \right) a^{2} \phi^{2}+(\phi')^{2}\right]\,,
  \\
  \frac{2\alpha'}{\alpha}&=\frac{a^{2}-1}{r}+8 \pi  r\left[\left(\frac{\omega^{2}}{\alpha^{2}}- \frac{\mu^2 (\sigma^2 - 2\phi^2)^2}{\sigma^4}\right) a^{2} \phi^{2}+(\phi')^{2}\right]\,,
  \\
  \phi''&=-\left\{1+a^{2}-8 \pi r^{2} \mu^2 a^{2}  \phi^{2} \frac{(\sigma^2-2\phi^2)^2}{\sigma^4} \right\} \frac{\phi'}{r}
  \\
&\quad{}-\left\{\frac{\omega^{2}}{\alpha^{2}}-\mu^{2} - \frac{4\mu^2\phi^2}{\sigma^4}(3\phi^2-2\sigma^2) \right\} \phi a^{2}\,,
\end{align*}
where primes stand for radial derivatives $\partial_r$. To obtain a physical solution, the following boundary conditions must be imposed on this system.
\begin{align}
\phi(0) &= \phi_0\,,\qquad \phi'(0) = 0\,,\qquad a(0)= 1\,,\label{bcPhiA}\\
\lim_{r\to\infty}\phi(r)&= 0\,,\quad  \lim_{r\to\infty} \alpha(r) a(r) = 1\,.\label{bcAlphaA}
\end{align}
$\phi_0$ can be specified arbitrarily and roughly determines the mass of the boson star. We can find a simpler system by rescaling the variables in the following manner,
\[
\tilde{\phi} \equiv \frac{\phi}{\sigma}, \quad \tilde{r} \equiv \mu r, \quad \tilde{t} \equiv \omega t, \quad \tilde{\alpha} \equiv(\mu / \omega) \alpha.
\]
Then the equations become
\begin{equation}
  \label{eq:BS-system}
\begin{aligned}
  &a'=\frac{a}{2}\left\{\frac{1-a^{2}}{\tilde{r}} + 8\pi\sigma^2\tilde{r}\left[\left(\frac{1}{\tilde{\alpha}^{2}}+ (1 - 2\tilde{\phi}^2)^2\right) a^{2} \tilde{\phi}^{2}+(\tilde{\phi}')^{2}\right]\right\}\,,\\
  &\tilde{\alpha}'=\frac{\tilde{\alpha}}{2}\left\{\frac{a^{2}-1}{\tilde{r}}+ 8\pi\sigma^2\tilde{r}\left[\left(\frac{1}{\tilde{\alpha}^{2}}- (1 - 2\tilde{\phi}^2)^2\right) a^{2} \tilde{\phi}^{2}+(\tilde{\phi}')^{2}\right]\right\}\,,\\
  &\tilde{\phi}''=-\left[1+a^{2}-8\pi \mu^2 \sigma^2 \tilde{r}^{2} a^{2} \tilde{\phi}^{2} (1-2\tilde{\phi}^2)^2  \right] \frac{\tilde{\phi}'}{\tilde{r}}\\
  &\qquad\qquad-\left[\frac{1}{\tilde{\alpha}^{2}} - 1 - 4\tilde{\phi}^2(3\tilde{\phi}^2 - 2)\right] \tilde{\phi} a^{2}\,,
\end{aligned}
\end{equation}
where primes now stand for derivatives with respect to $\tilde{r}$. To integrate these equations, we need to understand their asymptotic behavior. At the origin, $\tilde{r} = 0$, we can expand all quantities in a Taylor series to find
\begin{align*}
  a(\tilde{r})&=1 + \frac{4\pi\tilde{r}^2\sigma^2\tilde{\phi}_0^2}{3\tilde{\alpha}_0^2} \left[1 + \tilde{\alpha}_0^2 (1 - 2 \tilde{\phi}_0^2)^2\right] + \mathcal{O}(\tilde{r}^4)\,,\\
  \tilde{\alpha}(\tilde{r})&=\tilde{\alpha}_0 + \frac{4\pi \sigma^2\tilde{r}^2\tilde{\phi}_0^2}{3\tilde{\alpha}_0} \left[2 - \tilde{\alpha}_0^2 (1 - 2 \tilde{\phi}_0^2)^2\right] + \mathcal{O}(\tilde{r}^4)\,,\\
  \tilde{\phi}(\tilde{r})&= \tilde{\phi}_0 + \frac{\tilde{r}^2\tilde{\phi}_0}{6} \left[1-\frac{1}{\tilde{\alpha}_0^2} - 8 \tilde{\phi}_0^2 + 12\tilde{\phi}_0^4 \right] + \mathcal{O}(\tilde{r}^4)\,,
\end{align*}
where $\tilde{\phi}(0) = \tilde{\phi}_0$, $\tilde{\alpha}(0) = \tilde{\alpha}_0$. At large distances, the asymptotic behavior of $\phi$ is
\be
\tilde{\phi}(\tilde{r} \to \infty) \sim \frac{1}{\tilde{r}} \exp(-\tilde{r} \sqrt{1 - \tilde{\alpha}^{-2}}) \,.
\ee

Equation~(\ref{eq:potential}) results in the potential of mini boson stars in the limit $\sigma \to \infty$~\cite{Macedo:2013jja}. 
A SBS is not always dynamically stable against linear fluctuations. An unstable solution evolves on timescales possibly shorter than those of collision processes we aim to study, hence it is crucial to select linearly stable solutions as initial data.
The stability of spherically symmetric boson stars has been investigated with findings showing that stability changes at a mass extremum, meaning it is marginally stable at a particular value of $\phi_0$: $dM / d\phi_0 = 0$ (See II. C in Ref.~\cite{Siemonsen:2020hcg} for details). This result reveals, as indicated in Fig.~2 of Ref.~\cite{Collodel:2022jly}, that SBSs possess two stable and two unstable branches for any value of $\sigma \ll 1$ \cite{Boskovic:2021nfs,Cardoso:2021ehg,Kleihaus:2011sx,Tamaki:2011zza}.

The first stable branch of SBSs arises from the non-relativistic limit ($\phi_0 / \sigma \to 0$) with weak self-interactions, leading us to anticipate that it will yield results similar to those of the mini-boson star.
In contrast, the second stable branch is situated near $\phi_0 / \sigma$ and possesses significantly stronger self-interactions. This allows for a more compact configuration of SBSs compared to the first branch.

\begin{figure}[!htbp]
  \centering
  \includegraphics[width=0.45\textwidth]{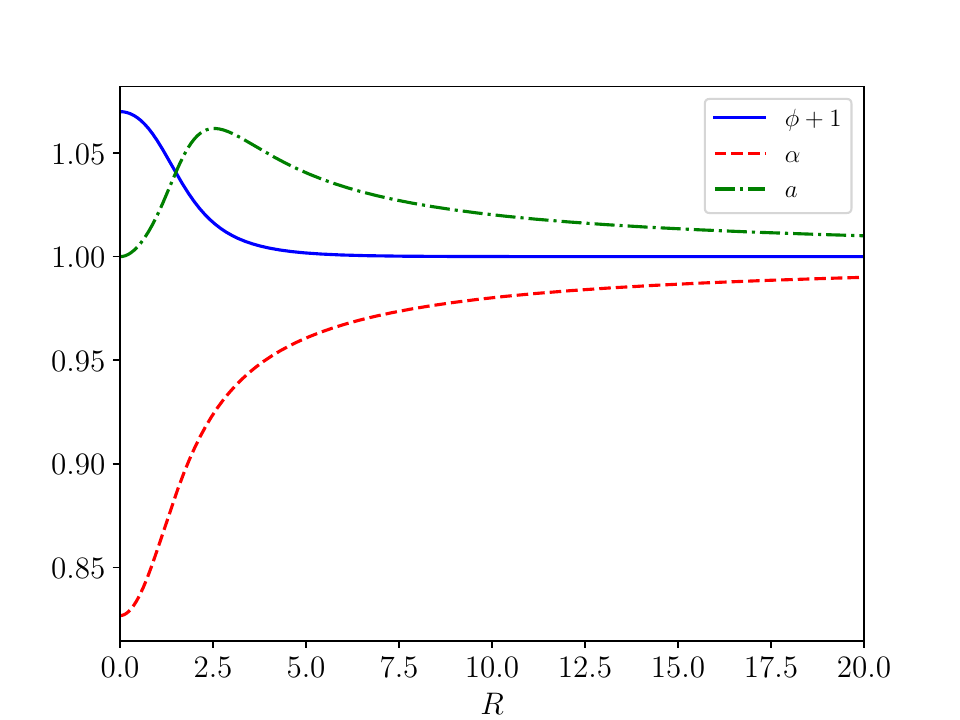}
  \caption{Scalar field and metric components as functions of the isotropic radial coordinate $R$ for an isolated SBS with mass $M=0.20$, $\tilde{\phi}_0 = 0.7,\,\sigma = 0.1,\,\omega=0.1,\, R_{98}=4.41$. In this work we focus on this specific SBS. \label{fig:solitonic_boson_star}}
\end{figure}

It is also worth noting that while SBSs may initially form in a dilute state, subsequent interactions and coalescence could lead to more compact configurations~\cite{Brito:2015yga,Boskovic:2021nfs}. Moving forward, we will standardize units such that $\mu = 1$. All our results will be shown and analyzed using this unit measure. In the following analysis, we will focus on the second branch and utilize a ground-state SBS characterized by $\tilde{\phi}_0 = 0.7$, $\sigma = 0.1$, $M = 0.20$, and $R_{98}=4.41$, where $R_{98}$ represents the radius that encompasses 98\% of the SBS mass. This configuration is depicted in Fig.~\ref{fig:solitonic_boson_star}.

\subsection{Black hole - boson star binary}

To construct initial data of BH-SBS spacetime, we transform radial coordinates of SBS into isotropic coordinates $R$ and superpose this solution with a boosted Schwarzschild BH, the details of which are demonstrated in Ref.~\cite{Cardoso:2022vpj}.

To evolve this system, we employ the Baumgarte-Shapiro-Shibata-Nakamura formulation of Einstein's equations~\cite{Nakamura87,Shibata:1995we,Baumgarte:1998te} for our numerical simulations and rely on the infrastructure of the Einstein Toolkit~\cite{Loffler:2011ay,Zilhao:2013hia,EinsteinToolkit:2022_11} for the numerical evolutions. Mesh refinement capabilities are facilitated by Carpet~\cite{Schnetter:2003rb}, apparent horizons are located and tracked using AHFinderDirect~\cite{Thornburg:1995cp,Thornburg:2003sf}, and BH mass is extracted using QuasiLocalMeasures~\cite{Dreyer:2002mx}. The spacetime metric and scalar field variables are evolved in time using the LeanBSSNMoL and ScalarEvolve codes~\cite{Canuda_zenodo_3565474,Cunha:2017wao}. We employ the method of lines, coupled with the fourth-order Runge-Kutta technique, to advance our equations over time. In the integration process, we use outgoing (radiative) boundary conditions alongside the common $1+\mathrm{log}$ and Gamma-driver gauge conditions \cite{Alcubierre08a}. For all simulations we use a square numerical domain with $x^i_\mathrm{min}=-430$, $x^i_\mathrm{max}=430$. We tested also with larger domain sizes and it did not change the final results. We consistently employ a minimum of $40$ points to cover the BH, thereby guaranteeing sufficient grid points to achieve satisfactory resolution and use the same grid structure as in Ref.~\cite{Cardoso:2022vpj}. We have run simulation \texttt{IVA} until $t \sim 1500$ and simulation \texttt{IVB} until $t \sim 800$. The data analysis for this project was performed utilizing the Python package ``kuibit''~\cite{kuibit}.

\subsection{Diagnostic tools}\label{sec:diagnostic_tools}

To gain a clearer understanding and more precise characterization of some of the physics involved, we track the following quantities, whose definitions can be found in Ref.~\cite{Cardoso:2022vpj}:
\begin{itemize}
  \item The spherical harmonics decomposition of the scalar field $\phi_{lm} (t, r)$ in the vicinity of the moving BH, using a frame that is comoving with the BH.
	We use a coordinate system where the dynamics is axi-symmetric, hence the only contributing multipoles have azimuthal number $m=0$.
  \item The energy $E^{\mathrm{rad}}$ and momentum $P^{\mathrm{rad}}$ radiated in GWs at large distances.
  \item The total energy density $Q_t$ of the scalar field into the BH horizon. %
\end{itemize}

\section{Numerical Results}
\begin{figure*}[tphb]
  \includegraphics[width=\linewidth]{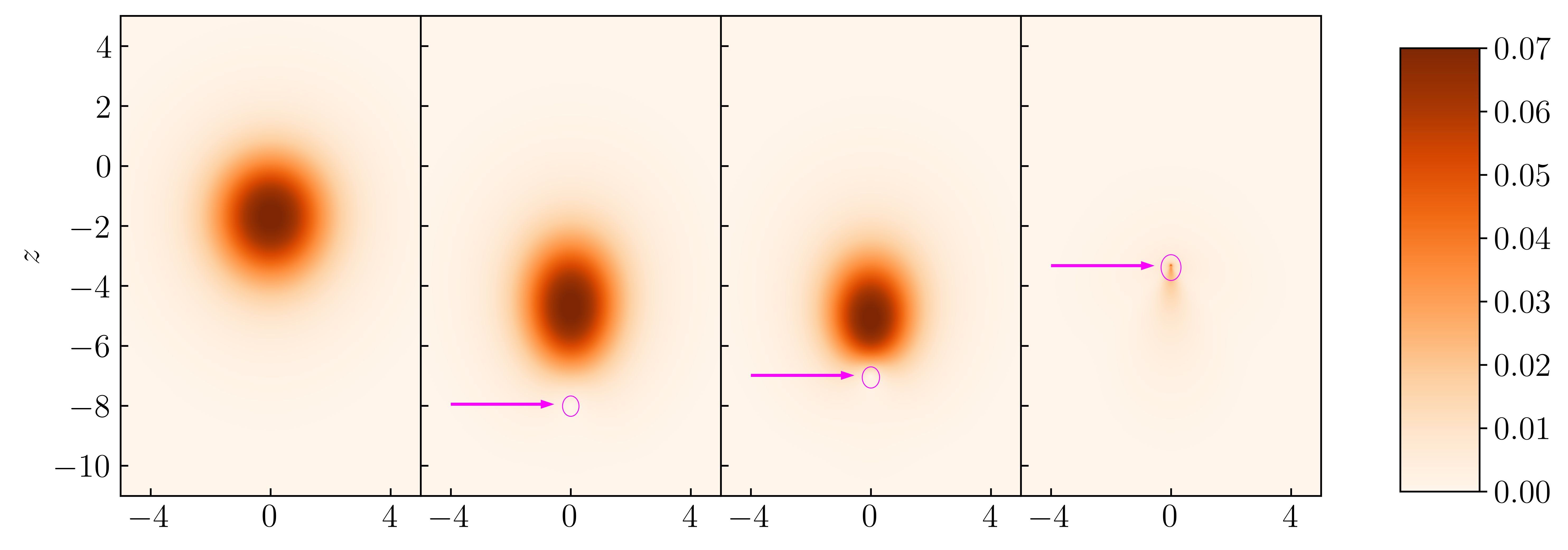}
  \caption{Snapshots of evolution for the simulation \texttt{IB}, where the BH and SBS are nearly of equal mass. Color intensity depicts scalar field absolute value $|\Phi|$. Snapshots are shown at instants $t=80\,,100\,,102\,,112$ from left to right. The pink lines depict contours of constant lapse function $\alpha = 0.2$, a rough measure for the location of the apparent horizon. This figure illustrates that the SBS undergoes considerable tidal distortion as it nears the BH, and ultimate near-total accretion by the BH.
\label{fig:snapshotsIB}}
\end{figure*}

\begin{figure*}[tphb]
  \includegraphics[width=\textwidth]{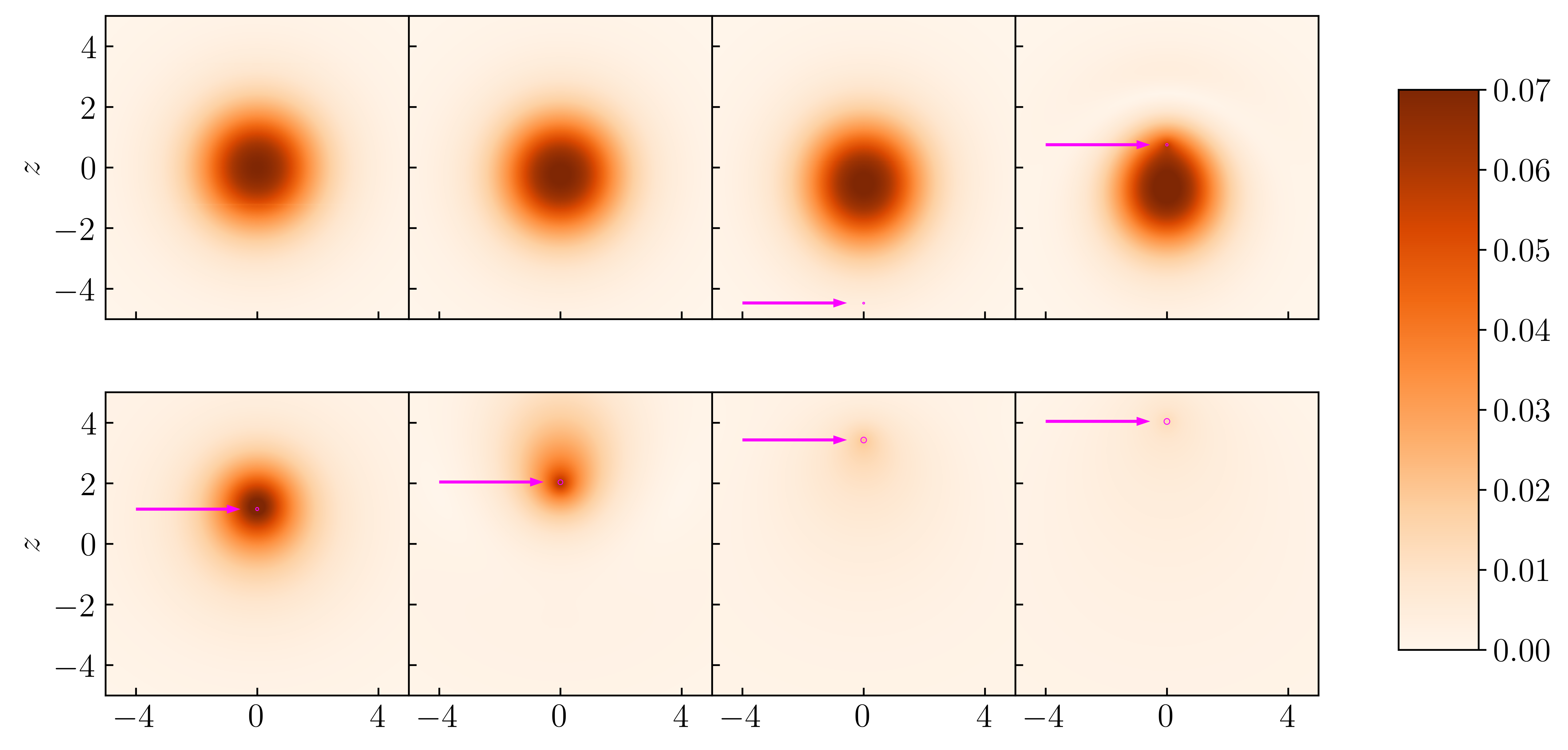}
  \caption{Snapshots of evolution, depicting the scalar field absolute value $|\Phi|$ for the simulation \texttt{IVB}. The top row displays snapshots taken at instants $t=0.0\,,79.36\,,94.72,$ and $107.52$ from left to right, while the bottom row shows snapshots taken at $t=120.32\,,130.56\,,140.8,$ and $145.92$ from left to right. As in the previous case, the pink lines depict contours of constant lapse function $\alpha = 0.2$, indicating the location of the apparent horizon. The pink circle in this figure is much smaller and harder to see compared to the one in Fig.~\ref{fig:snapshotsIB}, due to the significantly smaller size of the BH. In this figure, the SBS pulls back the BH during the collision process, as depicted in panels 5 and 6. Finally, the BH swallows the BS completely. Notice that when the BH first passes through the SBS, the tidal deformation of the SBS is quite inconspicuous. However, as the SBS accretes an increasing amount of the scalar field, the deformation becomes more pronounced.
    \label{fig:snapshotsIVB}}
\end{figure*}

\begin{table}[htb]
  \caption{List of simulations analyzed for collisions between a BH of mass parameter $M_{\mathrm{BH}}$ and an SBS with mass $M=0.20$. The BH is initially moving along the $z$-axis with a velocity of $v_0$ and starting from position $z_0 = -50$. The SBS is characterized by a frequency of $\omega=0.759$ and values of $\tilde{\phi}_0=0.7\,,\sigma = 0.1\,,\alpha_0=0.827$ at the origin. The total energy of the system, $M_{\mathrm{tot}}$, can be approximated using a Newtonian approach as $M_{\mathrm{tot}}=\Gamma M_{\mathrm{BH}}+M-\Gamma M_{\mathrm{BH}}M/z_0$, where $\Gamma$ is the Lorentz factor. The total momentum of the boosted BH is $\Gamma M_{\mathrm{BH}} v_0$. The simulations use a mass ratio of $q=M / M_{\mathrm{BH}}$ and a length ratio of ${\cal L}=R_{98} /(2M_{\mathrm{BH}})$ as parameters. It should be noted that initially the mass parameter $M_{\mathrm{BH}}$ is approximately equal to the irreducible mass $M_{\mathrm{irr}}$ to within 0.5\%. The irreducible mass can be calculated as $\mathcal{A} = 16 \pi M_{\mathrm{irr}}^2$, where $\mathcal{A}$ is the area of the apparent horizon. Recall that all results are presented in units where $\mu=1$. \label{table:simulations_parameters}}
  \begin{ruledtabular}
  \begin{tabular}{l@{\hskip 0.1in}c@{\hskip 0.1in}c@{\hskip 0.1in}c@{\hskip 0.1in}c@{\hskip 0.1in}c@{\hskip 0.1in}c@{\hskip 0.1in}} %
  Run          & $M_{\rm BH}$&${\cal L}$  & $v_0$   & $M_{\rm tot}$ &$P_{\rm tot}$ %
  \\
  \hline
  \texttt{IA} & 0.5         &4        & $10^{-4}$   &0.70           & 0             %
  \\
  \texttt{IB} & 0.5         &4        & $0.5$  &0.78           & 0.289        %
  \\
  \texttt{IIA} & 0.25        &9        & $10^{-4}$    &0.45           & 0            %
  \\
  \texttt{IIB} & 0.25        &9        & $0.5$    &0.49           & 0.144        %
  \\
  \texttt{IIIA} & 0.125        &18     & $10^{-4}$   &0.33           & 0            %
  \\
  \texttt{IIIB} & 0.125        &18     & $0.5$       &0.35           & 0.072        %
  \\
  \texttt{IVA} & 0.0625        &35     & $10^{-4}$  &0.26           & 0        %
  \\
  \texttt{IVB} & 0.0625        &35     & $0.5$       &0.27           & 0.036        %
  \end{tabular}
  \end{ruledtabular}
\end{table}
We conducted a study on a range of initial conditions, varying the initial mass and velocity of the BH. We use coordinates such that the SBS is initially at rest at the origin,
and the BH is located along the $z$ axis, initially at $(0, 0, z_0)$ and moving in the positive $z$-direction. Appendix~\ref{app:convergence} demonstrates the convergence of our numerical simulations.
The initial conditions are summarized in Table~\ref{table:simulations_parameters} and
our numerical results and findings are summarized in Table~\ref{table:simulations_results} and Figs.~\ref{fig:snapshotsIB}--\ref{fig:bound_state}. In the following subsections, we focus on two typical cases, Run \texttt{IIIB} and Run \texttt{IVB}, as our main interest lies in small BHs. However, to more clearly illustrate tidal deformation, we opt to showcase Run \texttt{IB} rather than Run \texttt{IIIB} in Section~\ref{sec:dynamics}.

\begin{table*}[tb]
  \caption{Summary of the results of the dynamical evolution of the initial data in Table~\ref{table:simulations_parameters}. Here, $M_f$ represents the final BH irreducible mass, and $v_f$ denotes the final BH velocity, calculated from the puncture trajectory. In parentheses, we display the expected value $M_{\rm BH} v_0/M_{\rm tot}$ based on momentum conservation, assuming that the entire BS is accreted onto the BH (note the strong agreement between these two estimates). $E^{\rm rad}$ and $P^{\rm rad}$ stand for the energy and momentum radiated in GWs, respectively. These values are calculated from $\psi_4$. Lastly, the total momentum and energy flux of the scalar field into the BH horizon are presented in the final two entries. Junk radiation is present in all cases, but its effect has been excluded. %
  \label{table:simulations_results}}
  \begin{ruledtabular}
  \begin{tabular}{l@{\hskip 0.1in}c@{\hskip 0.1in}c@{\hskip 0.1in}c@{\hskip 0.1in}c@{\hskip 0.1in}c@{\hskip 0.1in}c@{\hskip 0.1in}c@{\hskip 0.1in}c@{\hskip 0.1in}} %
  Run          & $M_{\rm BH}$&$M_f$ & $v_0$         & $v_f$                       & $10^4 E^{\rm rad}$ & $10^4 P_z^{\rm rad}$ &$Q_{t}^{\rm initial}$ & $Q_{t}^{\rm final}$\\
  \hline
  \texttt{IA} & $0.5$       &    $0.70$  & $10^{-4}$     & $-3 \times 10^{-3}$ (0) & $2.10$     & $0.26$ & $0.20$ & $2.6 \times 10^{-4}$ \\
  \texttt{IB} & $0.5$       &    $0.75$  & $0.5$         & $0.36 (0.36)$                &   $4.99$   & $-1.72$ & $0.20$ & $6.3 \times 10^{-4}$ \\
  \texttt{IIA} & $0.25$      &    $0.45$  & $10^{-4}$     & $-4.8 \times 10^{-3}$ (0) & $1.05$  & $0.08$ & $0.19$ & $2.0 \times 10^{-3}$ \\
  \texttt{IIB} & $0.25$      &    $0.48$  & $0.5$         & $0.27 (0.26)$          &  $3.09$    &  $-0.84$ & $0.20$ & $2.0 \times 10^{-3}$ \\
  \texttt{IIIA} & $0.125$       & $0.29$  & $10^{-4}$     & $-1.9\times 10^{-2}$ (0) & $0.46$   & $9.7 \times 10^{-3}$ & $0.19$ & $2.8 \times 10^{-2}$ \\
  \texttt{IIIB} & $0.125$   &    $0.31$     & $0.5$   &  $0.17 (0.18)$    &  $1.35$     &  $-0.31$ & $0.20$ & $2.4 \times 10^{-2}$ \\
  \texttt{IVA} & $0.0625$   &    $0.24$     & $10^{-4}$   &  $1.0 \times 10^{-3}(0)$   &  $0.29$     &  $0.005$ & $0.19$ & $1.9 \times 10^{-2}$ \\
  \texttt{IVB} & $0.0625$   &    $0.25$     & $0.5$             &  $0.12 (0.12)$   &  $0.52$     &  $-0.09$ & $0.20$ & $2.9 \times 10^{-2}$ \\
  \end{tabular}
  \end{ruledtabular}
\end{table*}


\subsection{Dynamics and accretion during collision}\label{sec:dynamics}

\begin{figure}[htbp]
  \centering
  \includegraphics[width=0.45\textwidth]{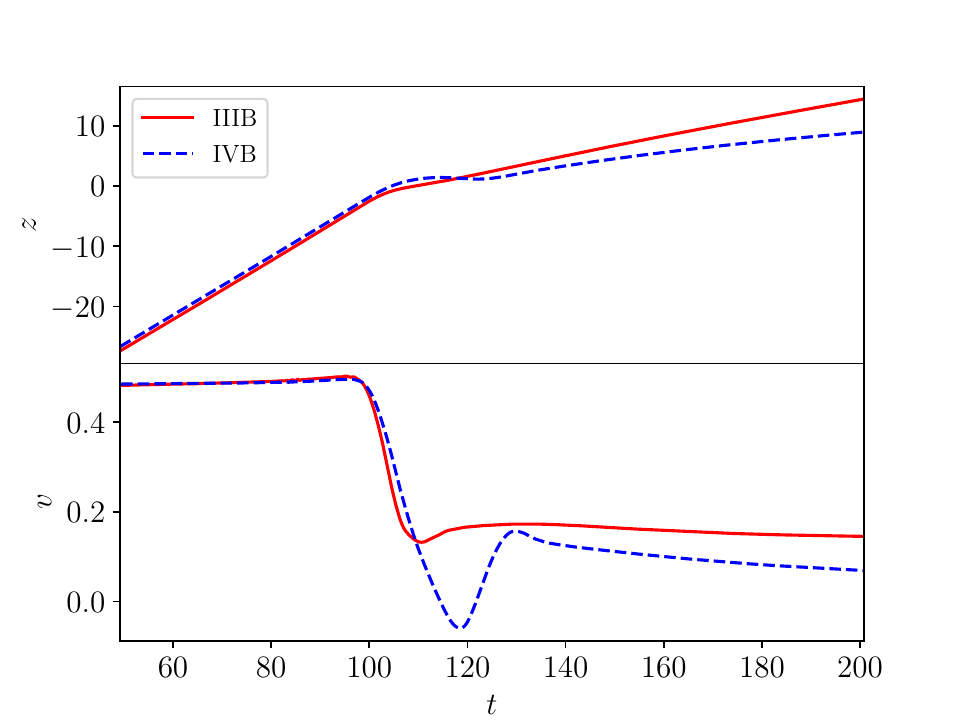}
  \caption{The puncture location $z$ and the velocity $v$ of the BH for simulations \texttt{IIIB} and \texttt{IVB}. They provide good estimates for the location and velocity of the BH, and these results demonstrate a clear interaction between the BH and the BS. Notably, in simulation \texttt{IVB}, the BH velocity becomes negative for a brief period as the BH is tidally captured by the SBS.\label{fig:bh_loc_v}}
\end{figure}

Snapshots of the evolution of the scalar field for initial data \texttt{IB} and \texttt{IVB} are shown in Figs.~\ref{fig:snapshotsIB} and \ref{fig:snapshotsIVB}, respectively. 
In both figures the tidal distortion of the boson star as the BH approaches is clear, probably due to their large (and positive) tidal Love numbers compared to compact systems~\cite{Mendes:2016vdr,Cardoso:2017cfl}. The distortion becomes more visible as the BH approaches the BS along the BH-BS axis. Tidal effects become crucial to capture the BH, and this is evident for simulation \texttt{IVB}: a much smaller BH, moving at half the speed of light is still captured by the SBS via tidal effects, ending up by accreting almost all of the SBS. The tidal capture is clearly illustrated in bottom panel of Fig.~\ref{fig:bh_loc_v}, where the BH's velocity even becomes negative for a short period.

Some of the main numerical results are reported Table~\ref{table:simulations_results}, which shows a few interesting aspects of this process. For all simulations we performed, across the different mass ratios, the BH ends up accreting the SBS. The reason for this is most likely three-fold: accretion and dynamical friction slows the BH down as the plunges through the SBS material~\cite{Annulli:2020lyc,Traykova:2021dua,Vicente:2022ivh}, but for the process to be fully effective, tidal capture ensures that the BH remains inside the SBS, eventually accreting it all or almost all. Accordingly, the velocity of the BH at late times is well estimated by simple momentum conservation as can be seen from Table~\ref{table:simulations_results}.

Given the velocity dependence of dynamical friction, it is unlikely that yet higher velocities would allow for the BH to cross the SBS and exit without first accreting it~\cite{Annulli:2020lyc,Traykova:2021dua,Vicente:2022ivh}, unless of the course one gets to more extreme mass ratios.
In fact, the BH absorption cross-section is the main factor that determines whether a BH can pass through a boson star without destroying it, and to decrease it one needs to make the BH smaller. Table~\ref{table:simulations_results} seems to indicate indeed that the residual scalar field increases for smaller BH mass. We can infer that for yet smaller BHs than those simulated here, the BH may pierce through the SBS, consuming only a small portion of scalar field, thereby leaving a smaller SBS in its wake. However, our numerical simulations would take a prohibitively long time to evolve such cases. %
As a result, within this theoretical framework, it is very difficult to verify whether a BH can pass through a boson star without destroying it.

\subsection{The tidal capture and gravitational-wave emission}

\begin{figure}[htbp]
  \centering
  \includegraphics[width=0.45\textwidth]{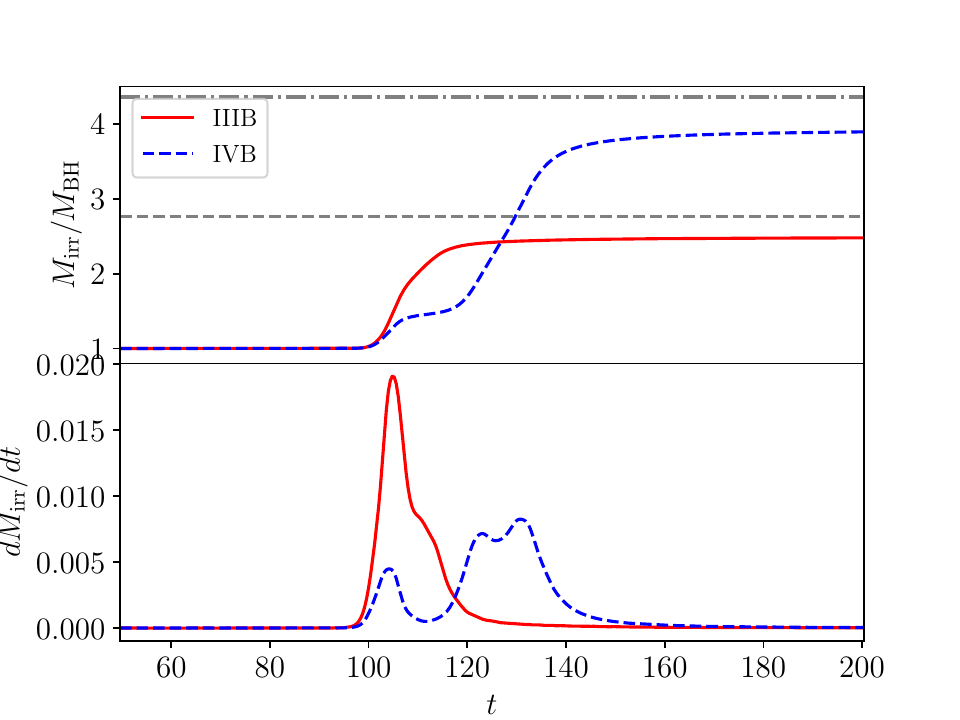}
  \caption{Accretion of scalar onto the BH. {\bf Top panel:} normalized BH irreducible mass $M_{\mathrm{irr}} / M_{\rm BH}$ for simulations \texttt{IIIB} and \texttt{IVB}. The gray lines are the normalized total mass $M_{\mathrm{tot}} / M_{\rm BH}$ given in Table~\ref{table:simulations_parameters}.
  At late times the BH mass approaches $M_{\rm tot}$, thus the BH ends up accreting the entire BS. {\bf Bottom panel:} accretion rate for the two different initial data. It is worth mentioning that for simulation \texttt{IVB}, there are two distinct stages of accretion that we believe are caused by tidal effects.
  \label{fig:bh_mass}}
\end{figure}
\begin{figure}[htbp]
  \centering
  \includegraphics[width=0.45\textwidth]{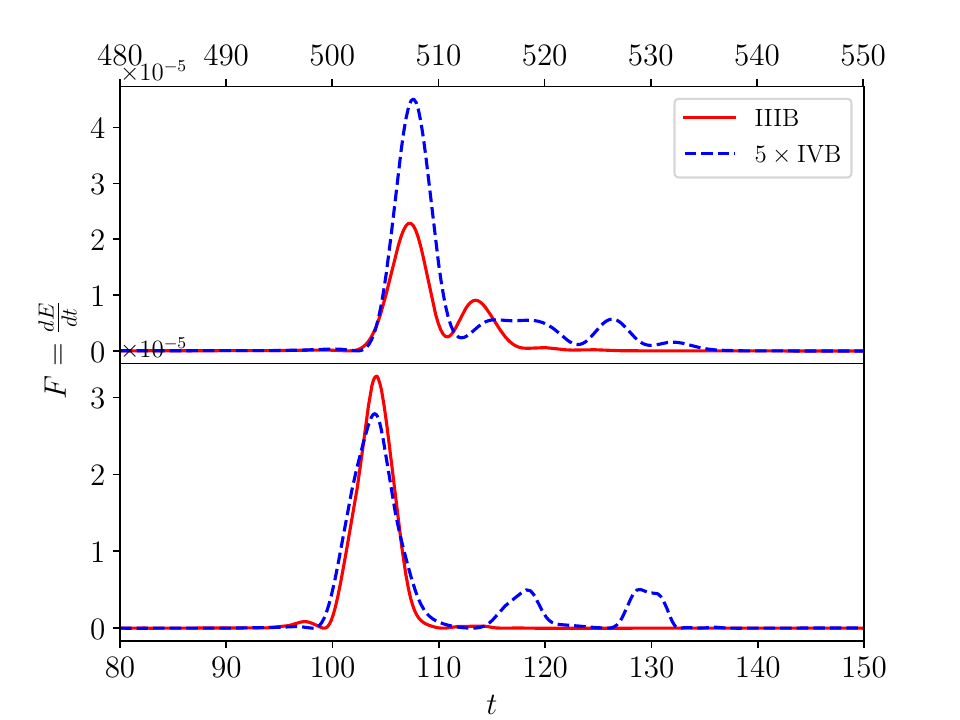}
  \caption{Energy flux $F = d E^{\mathrm{rad}} / dt$ of the GW for \texttt{IIIB} and \texttt{IVB}. \textbf{Top:} The energy flux obtained by integrating $\Psi_4$ over the sphere with radius $r = 400$. \textbf{Bottom:} The energy flux is calculated using the quadrupole approximation \eqref{eq:quadrupolar}, which requires numerical data from the simulation such as the puncture location, puncture velocity, and BH mass. \label{fig:gw_E}}
\end{figure}

When small BHs are tidally captured, we find that they oscillate around the center of the SBSs like a harmonic oscillator. The phenomenon is clearly observable for \texttt{IVB} case in both bottom panel of Fig.~\ref{fig:bh_mass} and upper panel of Fig.~\ref{fig:gw_E}, which features multiple peaks, indicating various stages of oscillation. The peaks depicted in the bottom panel of Fig.~\ref{fig:bh_mass} indicate a high accretion rate, suggesting that the BH is traversing the core of the SBS. Due to the deformation of the SBS, this core is identified as the region where the absolute value of the scalar field $|\Phi|$ reaches its maximum at this stage. Meanwhile, the peaks seen in the upper panel of Fig.~\ref{fig:gw_E} stem from the acceleration and deceleration of the relative movement between the SBS and the BH. It is worth noting that this oscillatory behavior is not clear in Fig.~\ref{fig:bh_loc_v}, given that the SBS have non-zero velocity in the lab frame.

To estimate the period of this oscillation and determine how close we are to the test particle threshold, we use the test particle approximation. By assuming this, we can use the lapse function $\alpha(R)$ of non-deformed SBS, which provides a good approximation of the gravitational potential. Therefore this function is also commonly referred to as the Newton potential. When considering a test particle oscillating within this potential, the acceleration it experiences can be determined from
\begin{equation}
\frac{d^2 R}{d t^2} = -\frac{d\alpha(r)}{dR} \,.
\end{equation}
The period given by the equation of motion is then
\begin{equation}
T = \frac{2\pi}{\sqrt{\alpha''(0)}} \approx 23.23 \,.
\end{equation}
Additionally, the period of the emitted gravitational wave corresponds to half of the motion period $T$, which for this case is $11.62$. This period is consistent with the first peak of gravitational waves observed in case \texttt{IVB}, as shown in Fig.~\ref{fig:gw_E}. This suggests that we are very near the test particle threshold for simulation \texttt{IVB}. Therefore, it is unlikely that we will observe any new phenomena at length ratios slightly larger but still within one order of magnitude. As we observe subsequent peaks, the oscillation period gradually decreases due to the BH accretion.

Our results for gravitational wave emission, derived using a fully relativistic approach, are presented in Table~\ref{table:simulations_results}. We have selected two typical cases to illustrate the waveform, as depicted in top panel of Fig.~\ref{fig:gw_E}. Following Ref.~\cite{Cardoso:2022vpj}, in the case that the BH mass is much smaller than SBS mass, the quadrupole approximation can be used to estimate the waveforms and radiated fluxes, with the BH moving along a spacetime geodesic defined by a radial position $r(t)$ in a background dictated by the SBS
\begin{equation}\label{eq:quadrupolar}
  \frac{d E}{d t}=\frac{8}{15} M_{\mathrm{BH}}^2(3 \dot{r} \ddot{r}+r \dddot{r})^2 .
\end{equation}
Nevertheless, the present scenario differs significantly from the Newtonian case, which is characterized by $\phi/\sigma \to 0$, rendering the methodology in Ref.~\cite{Cardoso:2022vpj} inapplicable in this context.  As an alternative solution, we use the puncture location, puncture veolcity and BH irreducible mass in numerical simulations instead of the original semi-analytic approximation. To reduce the impact of high frequency noise in numerical data of \texttt{IIIB} and \texttt{IVB}, we utilize a low-pass filter on both the puncture location and puncture velocity with a cutoff frequency $\omega_c = 2.5$ ($T = \frac{2\pi}{\omega_c} \approx 2.51$). From the numerical results, it can be seen that the dominant wavelength of the energy flux exceeds $2.51$, indicating that it would not significantly affect the main waveform. However, as depicted in bottom panel of Fig.~\ref{fig:gw_E}, this approximation fails to describe the peaks that follow the initial main peak in both \texttt{IIIB} and \texttt{IVB}, which emerge from the boson star, retaining only specific remnants during the final acceleration phase. This circumstance invalidates our initial assumption in the quadrupolar formula, where the BH mass $M_{\mathrm{BH}}$ is significantly smaller than the boson star mass and the trajectory of the BH is a geodesic on the SBS background. Ideally, substituting the puncture velocity with the relative velocity between black holes and SBS could lead to improved results. However, defining and calculating this relative velocity poses significant challenges. The result of quadrupolar approximation is given in bottom panel of Fig.~\ref{fig:gw_E}. However, in all instances, these values are too insignificant to exert any substantial impact on the system.

\begin{figure*}[thpb]
  \centering
  \subfloat[][]{\includegraphics[width=0.52\textwidth]{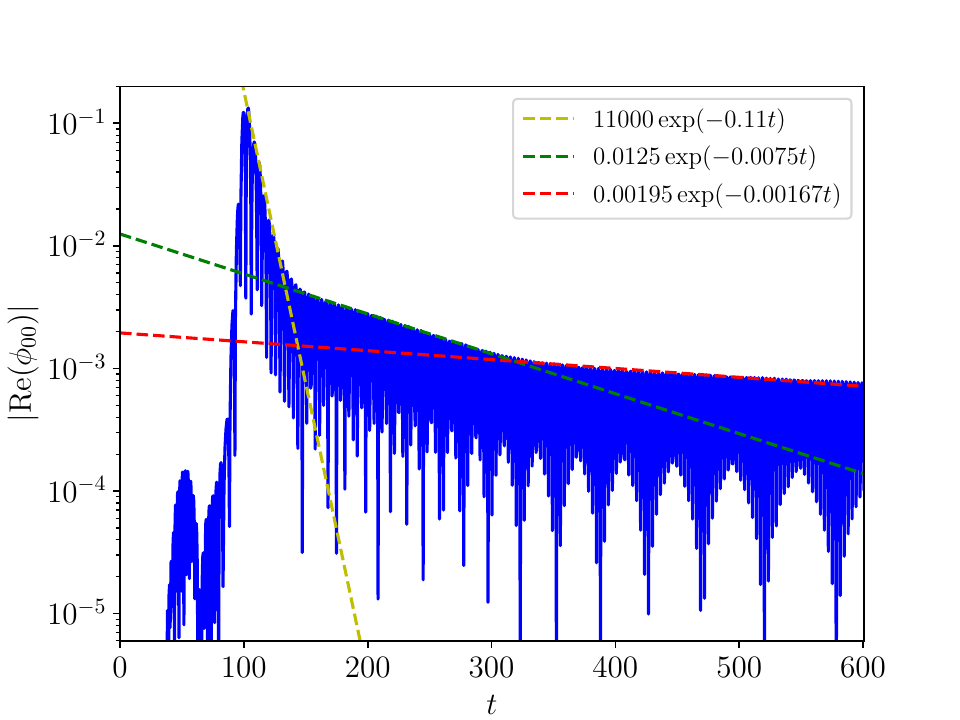}}
  \subfloat[][]{\includegraphics[width=0.52\textwidth]{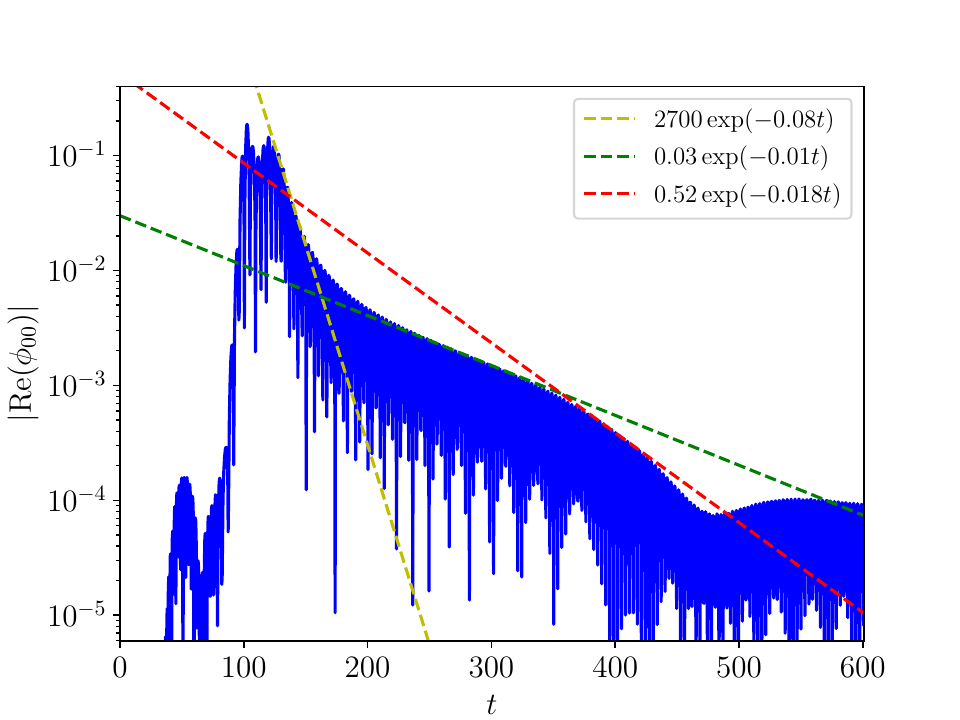}}
  \caption{Real part of the $l=m=0$ multipole of the scalar field on sphere $r = r_{\mathrm{BH}} + 1$ around the BH, where the left figure (a) is for \texttt{IIIB} and the right figure (b) is for \texttt{IVB}. The position of the sphere is taken as the position of the BH, and $r_{\mathrm{BH}}$ represents the radius of the BH horizon. As indicated by the dashed lines, it is evident that both \texttt{IIIB} and \texttt{IVB} display exponential decay, with varying rates at distinct stages. The rates of the dashed red lines are roughly consistent with the expectations from the quasi-bound state calculation using Leaver's method~\cite{Leaver:1985ax, 1986JMP....27.1238L, PhysRevD.34.384, Brito:2015oca}, which predicts $\omega_{\rm I} \simeq -0.0350,\,-0.00559,\, \mathbf{-0.00167},\,\ldots$ for \texttt{IIIB} ($M_{\rm f}=0.31$) and $\omega_{\rm I} \simeq \mathbf{-0.018},\, -0.0025,\, -0.00073,\,\ldots$ for \texttt{IVB} ($M_{\rm f}=0.25$), respectively. The modes are characterized by the principal quantum number $n$ \cite{Detweiler:1980uk}. It is notable that for \texttt{IIIB}, the corresponding mode is the $n = 3$ mode $\omega_{\rm I} \simeq -0.00167$, while for \texttt{IVB}, the corresponding mode is the $n = 1$ mode $\omega_{\rm I} \simeq -0.018$.
  This result indicate the existence of a ``gravitational atom''. The monopolar component experienced two growth phases. The decrease after the first growth is due to passing through the center of the SBS (defined as the place with the highest scalar field density). The second growth is caused by the BH pulling the SBS back and gradually swallowing it. Notice in panel (b) that the monopolar component has two highest peaks near the highest point, which implies that the BH oscillates at the center of the SBS.
	\label{fig:bound_state}}
\end{figure*}

\subsection{Late-time decay of the scalar}

As noted in Ref.~\cite{Cardoso:2022vpj} and demonstrated in the \texttt{IIIB} case in Fig.~\ref{fig:bh_mass}, when the mass of the BH is about half that of the BS, the BH enters a violent accretion phase, during which it absorbs most of the material from the BS. However, as shown in the \texttt{IVB} case in Fig.~\ref{fig:bh_mass}, the scenario deviates slightly. Here, the BH starts significantly smaller than the BS. After the BH is tidally captured and begins to increase in size, the subsequent stage of accretion becomes significantly more violent compared to the initial phase. In any case, a small portion of the BS remnants remains in a quasi-bound state, moving alongside the BH, which is typical for massive scalars. The small portion of BS remnants is expected to be mainly composed of spherical components and large wavelengths due to their lower accretion rate~\cite{Guzman:2012jc, Guzman:2004jw}. Therefore, we validate the quasi-bound states for $l = m = 0$ multiple, as predicted by perturbation theory, by using the spherical harmonic decomposition technique in the BH frame~\cite{Brito:2015oca}. Specifically, we calculate the quasi-bound state spectrum corresponding to the final black hole mass using Leaver's method \cite{Leaver:1985ax, 1986JMP....27.1238L, PhysRevD.34.384}. We find that the imaginary part of these modes align with the fitting result from one of the exponential decay stages. The results are shown in Fig.~\ref{fig:bound_state}. Note that while there are multiple exponential stages, it becomes challenging to compare them with the results of perturbation theory before the black hole mass reaches a stable stage, due to the variations in the black hole mass during the accretion process.

\section{Conclusion}

We have performed simulations involving BHs and SBSs, with length ratios as large as $\sim 35$.
Our objective is to investigate the interaction between bosonic structures with self-interaction, which could potentially represent dark matter, and BHs, as well as to determine the dynamical friction or accretion they induce on the BHs.
We find that the results are very similar to those obtained in Ref.~\cite{Cardoso:2022vpj}, even for a more compact SBS.
The presented results in this study, combined with those in with Ref.~\cite{Cardoso:2022vpj}, suggest that if a scalar field with self-interaction is a good model for describing dark matter, then the emergence of gravitational atoms will be very common in astrophysical environments, and thus possible to be detected by detectors~\cite{Nielsen:2019izz}.
As we expected, a gravitational atom comes into existence after collision, characterized by a massive BH surrounded by a quasi-bound state of the scalar field, known as the SBS remnant. This differs from the ones in Ref.~\cite{Guzman:2019gqc,Guzman:2022vxl}, which are primarily mixed-state solutions of the Schr{\"o}dinger-Poisson system including spherical and dipolar components.
Given that the oscillation period of the BH located at the center of the SBS is already close to the test particle limit, we do not expect any new phenomena to emerge until we reach extremely high length ratios, such as intermediate or even extreme mass ratios.
However, as we discussed in the introduction, the length ratio of solitonic cores formed by relativistic fuzzy dark matter and astrophysical black holes would be at least on the order of $10^6$, representing extreme mass ratio systems. Simulating such systems with our current computational infrastructure is not possible.
In all of our scenarios, we focus on a specific SBS, illustrated in Fig.~\ref{fig:solitonic_boson_star}. However, we anticipate that our findings will be similar for other SBS configurations with similar compactness to our present cases.

\begin{acknowledgments}
Z.Z.\ acknowledges financial support from China Scholarship Council (No.~202106040037).
V.C.\ is a Villum Investigator and a DNRF Chair, supported by VILLUM Foundation (grant no. VIL37766) and the DNRF Chair program (grant no. DNRF162) by the Danish National Research Foundation. V.C.\ acknowledges financial support provided under the European Union's H2020 ERC Advanced Grant ``Black holes: gravitational engines of discovery'' grant agreement
no.\ Gravitas--101052587. Views and opinions expressed are however those of the author only and do not necessarily reflect those of the European Union or the European Research Council. Neither the European Union nor the granting authority can be held responsible for them.
T.I.\ acknowledges financial support provided under the European Union's H2020 ERC, Starting
Grant agreement no.~DarkGRA--757480.
M.Z.\ acknowledges financial support by the Center for Research and Development in Mathematics and Applications (CIDMA) through the Portuguese Foundation for Science and Technology (FCT -- Fundação para a Ciência e a Tecnologia) -- references UIDB/04106/2020 and UIDP/04106/2020 -- as well as FCT projects 2022.00721.CEECIND, CERN/FIS-PAR/0027/2019, PTDC/FIS-AST/3041/2020, CERN/FIS-PAR/0024/2021 and 2022.04560.PTDC.
This work has further been supported by the European Horizon Europe staff exchange (SE) programme HORIZON-MSCA-2021-SE-01 Grant No.\ NewFunFiCO-101086251.
This project has received funding from the European Union's Horizon 2020 research and innovation programme under the Marie Sklodowska-Curie grant agreement No 101007855.
We thank FCT for financial support through Project~No.~UIDB/00099/2020.
We acknowledge financial support provided by FCT/Portugal through grants PTDC/MAT-APL/30043/2017 and PTDC/FIS-AST/7002/2020.
The results of this research have been achieved using the DECI resource Snellius based in The Netherlands at SURF with support from the PRACE aisbl, and
the Navigator cluster, operated by LCA-UCoimbra, through project~2021.09676.CPCA.

\end{acknowledgments}

\appendix

\section{Numerical convergence} \label{app:convergence}

We check the convergence of our numerical results by defining the usual convergence factor
\begin{equation}
  Q_{n}=\frac{f_{\Delta_{c}}-f_{\Delta_{m}}}{f_{\Delta_{m}}-f_{\Delta_{h}}}=\frac{\Delta_{c}^{n}-\Delta_{m}^{n}}{\Delta_{m}^{n}-\Delta_{h}^{n}}
\end{equation}
where $n$ denotes the order of the finite difference scheme employed, while $f_{\Delta_{c}}$, $f_{\Delta_{m}}$, and $f_{\Delta_{h}}$ represent the corresponding numerical solutions for a specified function $f$ at resolutions of $\Delta_c$, $\Delta_m$, and $\Delta_h$.

\begin{figure}[htbp]
  \centering
  \includegraphics[width=0.45\textwidth]{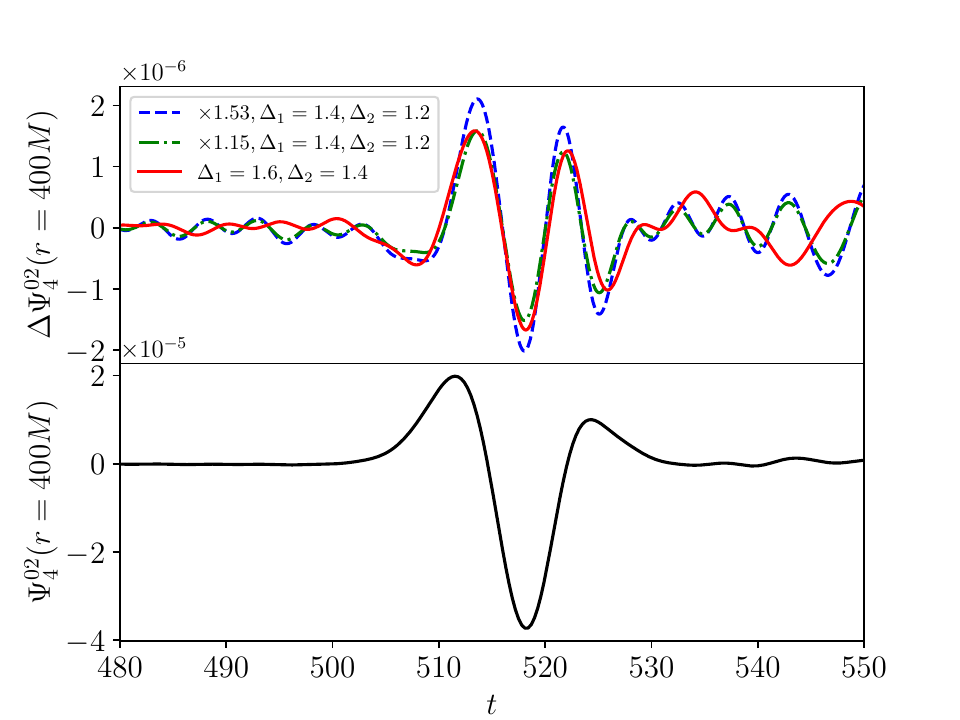}
  \caption{\textbf{Top:} The analysis of convergence for the $l = 0$, $m = 2$ multipole of $\Psi_4$, which was extracted at $r = 400M$, is presented. The blue line represents the expected result for a second-order convergence with a value of $Q_2 = 1.15$, while the green line illustrates the expected result for a fourth-order convergence, identified by $Q_4 = 1.53$. \textbf{Bottom:} The $l = 0$, $m = 2$ multipole of $\Psi_4$. The time interpolation order of Carpet is of the 2th order, whereas Multiple Thorn, which we employed to extract $\Psi_4^{02}$, utilizes a 3rd order interpolation order. Consequently, we anticipate that the convergence order of $\Psi_4^{02}$ will land between these two values—namely, the 2rd and 3th orders.\label{fig:cov}}
\end{figure}
We plot in Fig.~\ref{fig:cov} the convergence analysis for the $l = 0$, $m = 2$ multipole of $\Psi_4$, extracted at $r = 400M$, for configuration \texttt{IB}. The results are compatible with a convergence order between second and fourth order for physical waveform between $t = 500$ to $t = 530$.

\begin{figure}[htbp]
  \centering
  \includegraphics[width=0.45\textwidth]{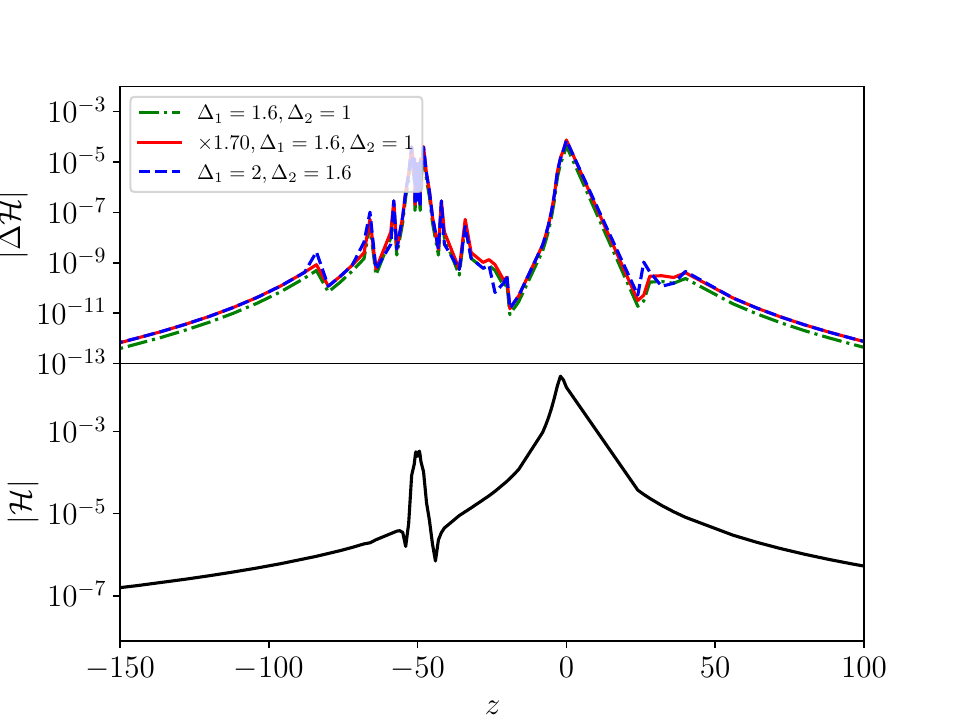}
  \caption{\textbf{Top:} Convergence of the Hamiltonian constraint violation at $t = 0$ for \texttt{IB}. The green line is multiplied by $Q_4 = 1.70$, the expected factor for fourth-order convergence. \textbf{Bottom:} We employ a technique known as \textit{Richardson extrapolation} to derive the value of the Hamiltonian constraint as $\Delta \to 0$.\label{fig:hc_cov_1b}}
\end{figure}
As illustrated in the bottom panel of Fig.~\ref{fig:hc_cov_1b}, the Hamiltonian constraint does not converge to zero, which is attributed to the superposition procedure used in constructing the initial data. To demonstrate the convergence that aligns with the finite difference scheme that was implemented, the top panel of Fig.~\ref{fig:hc_cov_1b} shows that the violation of the Hamiltonian constraint exhibits fourth-order convergence.

\begin{figure}[htbp]
  \centering
  \includegraphics[width=0.45\textwidth]{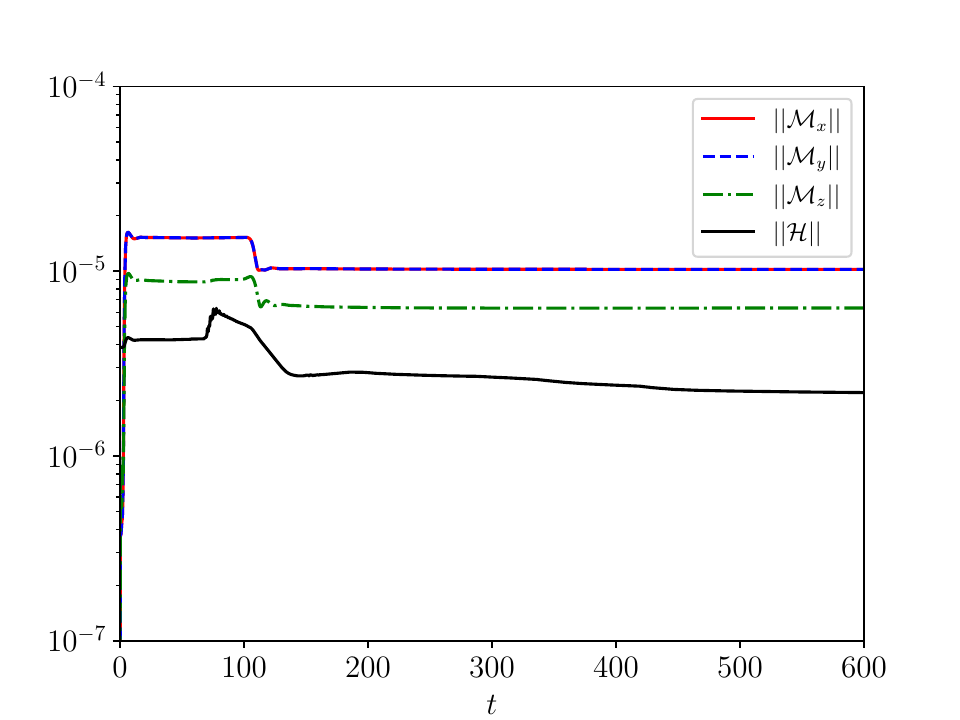}
  \caption{Violation of the Hamiltonian and momentum constraints as functions of time for run \texttt{IB}.\label{fig:cons_1b}}
\end{figure}
\begin{figure}[htbp]
  \centering
  \includegraphics[width=0.45\textwidth]{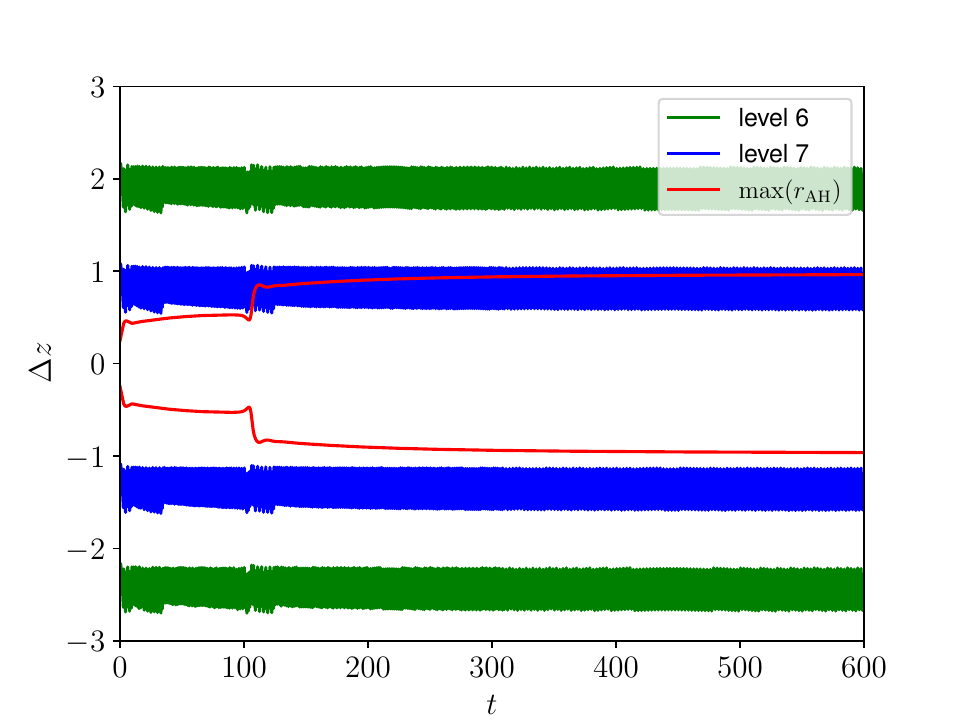}
  \caption{Difference in the $z$ direction between the boundary of the two finest refinement levels and the location of the BH puncture. The red line indicates the maximum apparent horizon radius of the BH. \label{fig:grid}}
\end{figure}

To ensure that constraint violations do not increase over time, we track the evolution of the $\ell^2$-norm of these violations, as shown in Fig.~\ref{fig:cons_1b}. Fig.~\ref{fig:grid} shows that the apparent horizon is consistently covered by finest level of the grid in Run \texttt{IB}, which demonstrates that Carpet tracks the grid structure effectively.

\bibliography{references}
\bibliographystyle{apsrev4-2}

\end{document}